# Estimating configurational entropy and energy of molecular systems from computed spectral density

**Jürgen Schlitter and Matthias Massarczyk**
Biophysics, Ruhr-University Bochum, 44780 Bochum, Germany
Correspondence: juergen.schlitter@rub.de

**Abstract**

Most methods for estimating configurational entropy from molecular simulation data yield upper limits except for harmonic systems where they are exact. Problems arise at diffusive systems and the presence of conformational transitions. Covariance-based methods, for instance, can considerably overestimate entropy when atomic positions change and cause large, but irrelevant spatial variances.

Here we propose a method called Spectrally Resolved Estimation (SRE) for entropy, energy and free energy which is based on the spectral density of vibrations and is inspired by the quasi-harmonic ansatz in solid-state physics. It (a) yields, in contrast with other methods, a lower limit of entropy and an upper limit of free energy, is (b) not corrupted by diffusion, (c) assigns entropic contributions to spectral features and thus reveals the localization and the type of motion underlying these contributions. The spatial extent of conformational transitions has no influence. The method applies to systems of finite volume and is exact for harmonic systems.

The exchange of solvent molecules is characteristic of biological macromolecules. Here SRE opens an avenue to the entropy and free energy of partially diffusive systems like proteins. It also enables studying channels, pumps, and enzymes which often contain several internal, functional water molecules that can swap position.

The assignment to particular motions can be done by normal modes or local mode analysis including visualization by band-pass filtering. Thus, SRE provides insight into the origin of configurational entropy and is expected to support the rational design of molecules from prodrugs up to engineered proteins.

This technical report demonstrates how thermodynamic quantities are gained and analyzed for small molecules by evaluation of spectra computed by quantum mechanical molecular dynamics simulation.

# Introduction

Configurational entropy is the entropy of a system which is at rest and free of rotation. For flexible macromolecules like proteins, it is the essential part of free energy which determines molecular structure and equilibria of compound binding and complex formation. A variety of methods (Baron et al., 2006; Suarez and Diaz, 2015) have been proposed and utilized for estimating configurational entropy from molecular simulation data and, with less accuracy, from experimental order parameters (Fleck et al., 2018) or mere structure (Goethe et al., 2018). Some approaches provide upper limits or bounds (King and Tidor, 2009; Schlitter, 1993). More recent methods for evaluating simulation trajectories like MIE (Killian et al., 2009), MIST (King and Tidor, 2009) and MCSA (Hensen et al., 2010) reach high accuracy by regarding higher correlations and anharmonic vibrations, which is demanding in terms of computing time.

Two problems still represent a particular challenge, firstly the exchange of solvent molecules in diffusive systems and secondly, the occurrence of conformational transitions. All methods which are based on spatial covariance overestimate entropy in these cases as atomic positions change largely and cause big, but irrelevant spatial variances (Chang et al., 2005). The use of force fluctuations (Ali et al., 2019; Hensen et al., 2014) or permutation reduction (Reinhard and Grubmuller, 2007) are possible approaches to treat molecular systems where problematic types of movements occur. An alternative way is offered by the quasi-harmonic ansatz in solid state physics (Kittel, 2004) which is not based on spatial fluctuations, but on spectral density of fluctuations. Density was recently also employed for calculating thermodynamic properties of infinite systems from solid state over liquid to gas using a two-phase model (Huang et al., 2011; Lin et al., 2003; Lin et al., 2010; Pascal et al., 2011).

Here we propose a method called Spectrally Resolved Estimation (SRE) which starts from the original quasi-harmonic ansatz(Kittel, 2004). It thus yields exact values for thermodynamic quantities of harmonically vibrating systems and furthermore applies as an approximation to systems which are restricted to a finite volume. It will be shown using statistical mechanics that SRE yields a lower limit to configurational entropy, which is information that was not available until now. The exchange of identical solvent molecules has no influence, for example, on calculated entropy as the spectrum remains unchanged. Switching between different conformations essentially results in average entropy because spectra are known to be averaged. This opens an avenue to the thermodynamics of macromolecules in solution. In particular, it enables studying proteins like channels and pumps which contain internal, functional water molecules that can swap position, and enzymes where entropy plays a role for the enhancement of reaction rates (Aqvist et al., 2017).

SRE was designed to resolve the entropy by attributing to each spectral band its contribution to total entropy. Using conventional techniques like standard normal mode analysis (NMA) and the recently developed local mode analysis (LMA) (Massarczyk et al., 2017), these spectral bands can be assigned to particular motions or groups of atoms and even be visualized by band-pass filtering of the trajectory, see supplement of (Massarczyk et al., 2017). Thus SRE gives insight into the origin of entropy and is expected to support the rational design of molecules from prodrugs up to engineered proteins.

This short technical report demonstrates how thermodynamic quantities are gained and analyzed for small molecules, starting from a quantum mechanically computed spectrum. We are aware that full validation will require application to macromolecular systems and comparison with other methods, which is beyond the scope of this introductory work.

## Theory

### Entropy and energy from spectral density

Let us assume that a system of N particles is in rest and free of rotations, but exerts harmonic vibrations. Spectral density defined as the number of oscillators per frequency interval $D(\omega) = dn/d\omega$ is here typical of a line spectrum,

$$D(\omega) = \sum_{i=1}^{3N-6} \delta(\omega - \omega_i)$$

where $\delta()$ is the delta distribution. Integration yields

$$(3N-6) = \int_0^\infty D(\omega)d\omega \qquad (1)$$

Each vibration contributes the molar entropy

$$s(\omega_i) = R\left\{\frac{\hbar\omega_i/k_B T}{\exp(\hbar\omega_i/k_B T) - 1} - \ln\left(1 - \exp(-\hbar\omega_i/k_B T)\right)\right\} \qquad (2)$$

of an harmonic oscillator (HO) where R denotes the gas constant, $\hbar$ the Planck constant, $k_B$ the Boltzmann constant, and T the absolute temperature. Total configurational entropy thus becomes

$$S_{conf} = \int_0^\infty s(\omega) D(\omega) d\omega \qquad (3)$$

in accordance with the expression used in solid state physics (Kittel, 2004). For application in molecular physics, a second expression is better suited which is obtained by combining (1) with (3) to

$$\widetilde{S} = (3N-6)\int_0^\infty s(\omega)D(\omega)d\omega / \int_0^\infty D(\omega)d\omega \qquad (4)$$

$\widetilde{S}$ represents entropy as a mean value $(3N-6)\langle s(\omega)\rangle_D$ obtained from a probability density that is proportional to spectral density $D(\omega)$. Eq. (4) is the relationship which we propose for estimating configurational entropy not only for harmonic systems but also for real systems at non-zero temperature which always have quasi-continuous spectra due to conformational dynamics and anharmonic interactions. Energy $\widetilde{U}$ and free energy $\widetilde{F}$ are further quantities that can be calculated analogously:

$$\widetilde{U} = (3N-6)RT \left\langle \frac{\hbar\omega/k_BT}{\exp(\hbar\omega/k_BT)-1} \right\rangle_D + E_0$$

$$\widetilde{F} = (3N-6)RT \left\langle \ln\left(1-\exp(-\hbar\omega/k_BT)\right) \right\rangle_D + E_0 \qquad (5)$$

$$E_0 = (3N-6)RT \left\langle \hbar\omega/2k_BT \right\rangle_D$$

$E_0$ is the zero-point energy. Thermodynamic quantities can be analyzed with respect to spectral contributions by calculating a running integral which, for the example of entropy, reads

$$s_{SR}(\omega) = (3N-6)\int_0^\omega s(\omega')D(\omega')d\omega' / \int_0^\infty D(\omega')d\omega' \xrightarrow{\omega \to \infty} \widetilde{S} \qquad (6)$$

This running integral allows connecting entropy with spectral bands and underlying motions. As the method can be applied to all thermodynamic quantities, it is denoted as Spectrally Resolved Estimation (SRE).

Spectral density can be calculated by the velocity autocorrelation function of mass-weighted velocities $\dot{q}_j$ of Cartesian coordinates or their Fourier transforms as

$$D(\omega) = \frac{2}{k_BT} \sum_{j=1}^{3N-6} d_j(\omega)$$

$$d_j(\omega) = \lim_{\tau\to\infty} \frac{1}{\tau} \left| \int_0^\tau \dot{q}_j(t)\exp(-i\omega t)dt \right|^2 \qquad (7)$$

The applicability of eqs. (2 – 6) depends on the behavior of density at $\omega = 0$. On the assumption that the system is confined to a volume inside a cubic box of length L one finds for each coordinate

$$d(0) = \lim_{\tau\to\infty} \frac{1}{\tau} \left| \int_0^\tau \dot{q}(t)dt \right|^2 = \lim_{\tau\to\infty} \frac{1}{\tau} |q(\tau)-q(0)|^2 = 0 \qquad (8)$$

This makes sure that $D(\omega)\ln(\omega)$ vanishes at $\omega = 0$ like at harmonic systems. In contrast, one finds $D(0) > 0$ at systems extending to infinity like bulk water where the quasi-harmonic ansatz has to be complemented by a gas-like term(Lin et al., 2003).

Spectral density $D(\omega)$ can also be derived from the intensity $F(\omega)$ of the (mechanical) fluctuation spectrum. This follows from the fact that each single oscillator with normal coordinate $Q_i$ and frequency $\omega_i$ delivers a spectral intensity $F_i(\omega) = \delta(\omega-\omega_i)\langle Q_i^2 \rangle = \delta(\omega-\omega_i)RT/\omega^2$. The second equation is a consequence of the equipartition theorem for the case that mass-weighted coordinates are used, which is always assumed here. The contribution to spectral density hence is $D_i(\omega) \equiv \delta(\omega-\omega_i) = F_i(\omega)\omega^2/RT$. More generally, this leads to the well-known relationship $D(\omega) = F(\omega)\omega^2/RT$ between spectral density and spectral intensity(Thomas et al., 2013).

By non-linear effects like conformational transitions and anharmonicity, molecular dynamics simulations at finite temperature yield continuous densities for which eqs. (2 – 6) are approximations. Their impact on thermodynamics is considered in the next section. On the technical side, the above expressions are explicitly proportional to the system size and spectral density needs to be determined only up to a constant factor, which facilitates the application and avoids numerical errors.

**Free energy and entropy as limits**

SRE is exact for harmonic systems, but is expected to return errors when applied to systems that are anharmonic because of a conformational equilibrium or anharmonicity of the potential which both also result in line broadening, i.e. deviation from the simple NMA line spectrum. The reason is that a system can take (at least) two thermodynamic states A and B with different spectra and the partition function (pf) hence is a sum

$$Z = Z_A + Z_B \tag{9}$$

This is immediately clear in case of a conformational equilibrium with different potential wells. For an anharmonic potential like the Morse potential, the system is at low temperature well described by the pf $Z_A$ of a harmonic system. At higher temperature, however, lower frequencies arise due to the anharmonic, wider potential at high energy. This effect can also be taken into account by an admixture $Z_B$ which dominates at high temperature and yields lower frequencies. The splitting into two states is the elementary, crucial step for line broadening and explains the limiting character of thermodynamic quantities calculated by SRE.

State A occurs with a probability $p_A = Z_A / Z$ and state B with $p_B = Z_B / Z$. We assume that each state has its own NMA spectrum and that probabilities are proportional to the spectral density which arises in state A and B, resp. Energy is calculated from the pf by taking the derivative $\partial_\beta$ with respect to $\beta = 1/k_B T$ and becomes

$$\begin{aligned} U &\equiv -\partial_\beta \ln Z \\ &= -p_A \partial_\beta \ln Z_A - p_B \partial_\beta \ln Z_B \\ &= p_A U_A + p_B U_B = \widetilde{U} \end{aligned} \tag{10}$$

Obviously, the weighted mean $\widetilde{U}$ yields the exact energy associated with $Z$.

For free energy we start from the inequality

$$Z \geq p_A Z_A + p_B Z_B \tag{11}$$

which follows from eq. (9) by taking the square of the pf. Then one finds

$$F \equiv -\beta^{-1} \ln Z$$
$$\leq -\beta^{-1} \ln \left( p_A Z_A + p_B Z_B \right)$$
$$\leq -\beta^{-1} \left( p_A \ln Z_A + p_B \ln Z_B \right) \quad (12)$$
$$= p_A F_A + p_B F_B = \widetilde{F}$$

Here the first inequality follows from eq. (11) together with the monotonous behavior of the logarithm, while the second one derives from the fact that the logarithm is a concave function. The weighted mean $\widetilde{F}$ overestimates the true free energy $F$.

Entropy thus obeys the inequality

$$S \equiv U/T - F/T$$
$$\geq \widetilde{U}/T - \widetilde{F}/T \quad (13)$$
$$= p_A S_A + p_B S_B = \widetilde{S}$$

with the consequence that the weighted mean $\widetilde{S}$ underestimates the true entropy $S$. In summary, when thermodynamic quantities are approximated as weighted means and weights are proportional to spectral density, then the calculated energy is correct, free energy is an upper limit and entropy a lower limit. Conformational transitions $A \to B$ of whatever size have no influence on the weighted means except by accompanying spectral changes. For harmonic systems where $Z_B = 0$, the method yields exact values.

## Application to benzene derivatives

The aromatic aminobenzene aniline $C_6H_7N$ is and metoxybenzene anisole $C_6H_7OCH_3$ are chosen to demonstrate the application of the method because the vibrational spectra are known and assigned both theoretically and experimentally and the gas phase standard entropy is available. Starting structures generated with Pymol (Schrödinger, 2015) were energy minimized and equilibrated over 10 ps using TeraChem (PetaChem, 2017) with camB3LYP 6-31G* and dispersion correction D3. Finally, 250 ps trajectories were calculated in production runs with time steps of 0.5 ps, Langevin thermostat with 0.2 ps time constant to keep temperature near 300 K. Center-of-mass motion and system rotation were suppressed.

Spectral intensity $F(\omega)$ was calculated from the trajectories of mass-weighted Cartesian coordinates $q_i(t) = \sqrt{m_i} x_i(t)$ after subtraction of the time average so that $\overline{q_i(t)} = 0$. The vector of these displacements, $\mathbf{q}(t) = \left( q_1(t)...q_{3N}(t) \right)$, is Fourier transformed to become $\hat{\mathbf{q}}(\omega) = \left( \hat{q}_1(\omega)...\hat{q}_{3N}(\omega) \right)$ and yields spectral intensity

$$F(\omega) = \hat{\mathbf{q}}(\omega) \cdot \hat{\mathbf{q}}^*(\omega) = \sum_{i=1}^{3N} \hat{q}_i(\omega) \hat{q}_i^*(\omega) \quad (14)$$

and spectral density

$$D(\omega) = const\ F(\omega)\omega^2.  \tag{15}$$

For the interpretation of spectra we go back to results of normal mode analysis (NMA) using density function theory for aniline (Wojciechowski et al., 2003) validated Raman and infrared (IR) spectroscopy. For anisole, resonance enhanced multiphoton ionization (REMPI) experiments (Hoffmann et al., 2006) provide complete, detailed spectra where all bands have been assigned to vibrational modes by comparison with the computed normal modes. Although NMA provides only unbroadened absorption lines at 0 K, they enable interpretation of our 300 K simulation results.

The SRE analysis is here performed for entropy which is of practical interest and can be compared with experimental standard values, see table 1. The workflow from spectral density to entropy is shown in Fig. 1.

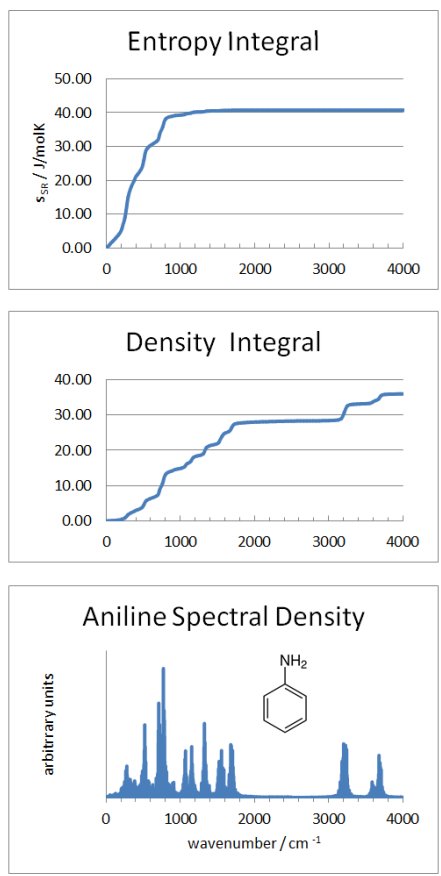

Figure 1. Workflow of the calculation of entropy $\tilde{S}$ for the example aniline. From spectral density (bottom) the running integral (middle) is calculated which here is normalized to a maximum value of 3N-6 = 36. Steps in the integral indicate the position of narrow bands. The entropy (top) is calculated as a running integral according to eq. (6) which converges to a final value representing the desired entropy $\tilde{S}$, here 40.7 J/molK. Hereafter, in the abscissa frequency is replaced by the more familiar wavenumber in cm$^{-1}$.

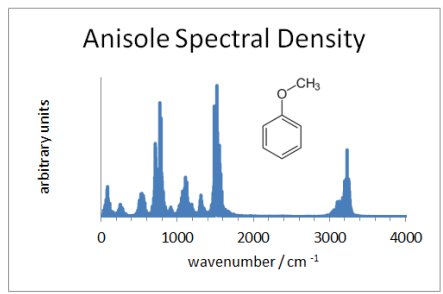

Figure 2. The spectral density of anisole shows a low-frequency peak near 80 cm$^{-1}$ of the OCH$_3$ group contributing about one third to entropy, and a strong band near 1400 cm$^{-1}$. The typical peaks of the aniline NH$_2$ group near 1700 and above 3500 cm$^{-1}$ occurring in Fig. 1 are missing.

The density integral indicates clearly that 7/36 of total spectral density is located above 2500 cm$^{-1}$ in agreement with NMA where 7 of 36 lines were found in that region. This provides an independent check of calculated spectral density. Vibrations above 2500 cm$^{-1}$ are hydrogen stretching modes with negligible entropy. Above 1000 cm$^{-1}$ most stretching and hydrogen bending modes are located which provide half of spectral density but contribute only 15 % to entropy (aniline), whereas the rest originates from the middle and far infrared region below 1000 cm$^{-1}$. Here out-of-plane bending of hydrogens and heavy-atom bond stretching in the benzene ring and infrared-active ring deformation modes predominate at 755 – 968 cm$^{-1}$ (aniline). Typical of aniline are the broad NH$_2$ torsion band around 300 cm$^{-1}$, a NH$_2$ wagging peak near 540 cm$^{-1}$ and a strong band around 1700 cm$^{-1}$ which is missing in anisole and apparently indicates scissoring of the amino group.

For comparison, Fig. 2 shows the spectrum of anisole which is similar to aniline between 600 and 1500 cm$^{-1}$, but shows features that are typical of the metoxy group in other regions: a C-OCH$_3$ torsion peak is seen at 80 cm$^{-1}$ and a O-CH$_3$ torsion peak at 250 cm$^{-1}$, whereas the COC ether bending mode expected near to 433 cm$^{-1}$ (from NMA) makes no visible contribution. The low-lying torsion peak of NMA is shifted from 90 to 80 in accordance with the experiment and gives an idea of line broadening in comparison to NMA. The entropy of anisole is larger mainly due to the low-frequency torsion of the massive metoxy group which contributes about one third to entropy.

Table 1. Entropy $\tilde{S}$ calculated from spectral density using eq. (4) with scaling factors 0.98 ± 0.01. Experimental standard entropy in the gas phase at 298 K yields configurational entropy $S^0_{conf}$ by subtracting entropies of translation and rotation obtained by ORCA and corrected by the symmetry number σ. For comparison, also the vibrational entropy is shown which results from the vibrational NMA frequencies computed and scaled by the authors.

|  | Aniline (Aminobenzene) | Anisole (Metoxybenzene) |
|---|---|---|
| $S^0$ / (J/molK) | 317.9[a] | 342.0[a] ; 360.6[b] |
| $S^0_{trans}$ | 165.4[c] | 167.3[c] |
| $S^0_{rot}$ | 112.8[c] | 117.0[c] |
| σ | 2 | 2 |
| **$S^0_{conf}$** | **45.5** | **72.8 ± 9.3** |
| **$\tilde{S}$** | **42.2 ± 0.8** | **63.7 ± 0.9** |
|  |  |  |
| $S^{NMA}_{vib}$ | 40.3[d] | 61.8[e] |

a (Stull, 1969), b (Kawaki et al., 1988), c ORCA (Neese, 2018), d (Wojciechowski et al., 2003), e (Hoffmann et al., 2006)

The results for entropy are comprised in table 1. Entropy was calculated for a range of frequency scaling factors which cannot be given *a priori* for the dynamical simulation. As expected, this method based on the vibrational spectrum of a dynamics simulation yields values $\tilde{S}$ below the configurational entropy $S^0_{conf}$ derived from experiments. Also normal mode analysis provides a lower limit here, but is restricted to small molecules with quasiharmonic dynamics.

Table 2. Energy and free energy from spectral density using eq. (5) with scaling factors 0.98 ± 0.01.

|  | Aniline (Aminobenzene) | Anisole (Metoxybenzene) |
|---|---|---|
| $\widetilde{U} / (kJ/mol)$ | 327.6 ± 3.4 | 357.9 ± 3.8 |
| $\widetilde{F}$ | 314.9 ± 3.3 | 338.8 ± 3.7 |
| $\Delta U = \widetilde{U} - E_0$ | 8.4 ± 3.4 | 10.4 ± 3.8 |
| $\Delta F = \widetilde{F} - E_0$ | -4.4 ± 0.1 | -8.7 ± 0.1 |

Table 2 presents energy and free energy for both molecules which here reflect the effect of the change in the side group. $\widetilde{U}$ and $\widetilde{F}$ measure the distance from the electronic ground state of the optimized structure. The differences $\Delta\widetilde{U}$ and $\Delta\widetilde{F}$ give the distance from the molecular vibrational ground state.

## Conclusion

In summary, SRE has proven to be an easily applicable method for computing and analyzing thermodynamic quantities of finite systems. In the examples, it delivers an approximation to configurational entropy which is lower than the value derived from experiments. This is in accordance with the presented theory which also predicts that free energy is overestimated while energy is well reproduced. The method is solely based on the unnormalized spectral density of the molecular system under consideration and gives insight into the origin of entropy by connecting contributions with spectral peaks or bands which can be localized by comparison with NMA (as done here) or using LMA (Massarczyk et al., 2017) for localizing and visualizing bands by band-pass filtering of the trajectory. The computing time is determined by the calculation of the trajectory which at least for small molecules has to be gained by quantum mechanical methods, while fast Fourier transform and SRE itself are negligible. The statistical accuracy which was not in the focus of this study will depend on the length and quality of the trajectory. For a small molecule already, entropy is dominated by low-frequent modes in the middle and far infrared region. At large molecules, molecular mechanics simulations will be required which reproduce reliably the low-frequency (terahertz) domain.

# References

Ali, H.S., Higham, J. and Henchman, R.H. (2019) Entropy of Simulated Liquids Using Multiscale Cell Correlation. *Entropy*, 21.

Aqvist, J., Kazemi, M., Isaksen, G.V. and Brandsdal, B.O. (2017) Entropy and Enzyme Catalysis. *Accounts of Chemical Research*, 50, 199-207.

Baron, R., van Gunsteren, W.F. and Hunenberger, P.H. (2006) Estimating the configurational entropy from molecular dynamics simulations: anharmonicity and correlation corrections to the quasi-harmonic approximation. *Trends in Physical Chemistry*, 87-122.

Chang, C.E., Chen, W. and Gilson, M.K. (2005) Evaluating the accuracy of the quasiharmonic approximation. *Journal of Chemical Theory and Computation*, 1, 1017-1028.

Fleck, M., Polyansky, A.A. and Zagrovic, B. (2018) Self-Consistent Framework Connecting Experimental Proxies of Protein Dynamics with Configurational Entropy. *Journal of Chemical Theory and Computation*, 14, 3796-3810.

Goethe, M., Gleixner, J., Fita, I. and Rubi, J.M. (2018) Prediction of Protein Configurational Entropy (Popcoen). *Journal of Chemical Theory and Computation*, 14, 1811-1819.

Hensen, U., Grater, F. and Henchman, R.H. (2014) Macromolecular Entropy Can Be Accurately Computed from Force. *Journal of Chemical Theory and Computation*, 10, 4777-4781.

Hensen, U., Lange, O.F. and Grubmuller, H. (2010) Estimating Absolute Configurational Entropies of Macromolecules: The Minimally Coupled Subspace Approach. *Plos One*, 5.

Hoffmann, L.J.H., Marquardt, S., Gemechu, A.S. and Baumgartel, H. (2006) The absorption spectra of anisole-h8, anisole-d3 and anisole-d8. The assignment of fundamental vibrations in the S-0 and the S-1 states. *Physical Chemistry Chemical Physics*, 8, 2360-2377.

Huang, S.N., Pascal, T.A., Goddard, W.A., Maiti, P.K. and Lin, S.T. (2011) Absolute Entropy and Energy of Carbon Dioxide Using the Two-Phase Thermodynamic Model. *Journal of Chemical Theory and Computation*, 7, 1893-1901.

Kawaki, H., Masuda, F. and Sasaki, Y. (1988) Substituent Entropy Constant Sigma-S-Degrees and Descriptor Mu-2-Alpha of Sym-Disubstituted Benzene-Derivatives - Estimation and Validity of Novel Quantitative Structure Activity Relationships Descriptors. *Chemical & Pharmaceutical Bulletin*, 36, 4814-4820.

Killian, B.J., Kravitz, J.Y., Somani, S., Dasgupta, P., Pang, Y.P. and Gilson, M.K. (2009) Configurational Entropy in Protein-Peptide Binding: Computational Study of Tsg101 Ubiquitin E2 Variant Domain with an HIV-Derived PTAP Nonapeptide. *Journal of Molecular Biology*, 389, 315-335.

King, B.M. and Tidor, B. (2009) MIST: Maximum Information Spanning Trees for dimension reduction of biological data sets. *Bioinformatics*, 25, 1165-1172.

Kittel, C. (2004) *Introduction to Solid State Physics*. Wiley John + Sons.

Lin, S.T., Blanco, M. and Goddard, W.A. (2003) The two-phase model for calculating thermodynamic properties of liquids from molecular dynamics: Validation for the phase diagram of Lennard-Jones fluids. *Journal of Chemical Physics*, 119, 11792-11805.

Lin, S.T., Maiti, P.K. and Goddard, W.A. (2010) Two-Phase Thermodynamic Model for Efficient and Accurate Absolute Entropy of Water from Molecular Dynamics Simulations. *Journal of Physical Chemistry B*, 114, 8191-8198.

Massarczyk, M., Rudack, T., Schlitter, J., Kuhne, J., Kotting, C. and Gerwert, K. (2017) Local Mode Analysis: Decoding IR Spectra by Visualizing Molecular Details. *Journal of Physical Chemistry B*, 121, 3483-3492.

Neese, F. (2018) Software update: the ORCA program system, version 4.0. *Wiley Interdisciplinary Reviews-Computational Molecular Science*, 8.

Pascal, T.A., Lin, S.T. and Goddard, W.A. (2011) Thermodynamics of liquids: standard molar entropies and heat capacities of common solvents from 2PT molecular dynamics. *Physical Chemistry Chemical Physics*, 13, 169-181.

PetaChem, L. (2017) *User's GuideTeraChem v1.9*, 26040 Elena RoadLos Altos Hills, CA 94022.

Reinhard, F. and Grubmuller, H. (2007) Estimation of absolute solvent and solvation shell entropies via permutation reduction. *Journal of Chemical Physics*, 126.

Schlitter, J. (1993) Estimation of Absolute and Relative Entropies of Macromolecules Using the Covariance-Matrix. *Chemical Physics Letters*, 215, 617-621.

Schrödinger, L. (2015) *The PyMOL Molecular Graphics System, Version 1.8*.

Stull, D.R. (1969) *The chemical thermodynamics of organic compounds*. Wiley, New York.

Suarez, D. and Diaz, N. (2015) Direct methods for computing single-molecule entropies from molecular simulations. *Wiley Interdisciplinary Reviews-Computational Molecular Science*, 5, 1-26.

Thomas, M., Brehm, M., Fligg, R., Vohringer, P. and Kirchner, B. (2013) Computing vibrational spectra from ab initio molecular dynamics. *Physical Chemistry Chemical Physics*, 15, 6608-6622.

Wojciechowski, P.M., Zierkiewicz, W., Michalska, D. and Hobza, P. (2003) Electronic structures, vibrational spectra, and revised assignment of aniline and its radical cation: Theoretical study. *Journal of Chemical Physics*, 118, 10900-10911.